\documentclass[lettersize,journal]{IEEEtran}
\usepackage{amsmath,amsfonts}
\usepackage[noend]{algpseudocode}
\usepackage{algorithm}
\usepackage{array}
\usepackage[caption=false,font=normalsize,labelfont=sf,textfont=sf]{subfig}
\usepackage{textcomp}
\usepackage{stfloats}
\usepackage{url}
\usepackage{verbatim}
\usepackage{graphicx}
\usepackage{cite}
\usepackage{amssymb,amsthm}
\usepackage{blkarray}
\usepackage{enumerate}

\newtheorem{definition}{Definition}
\newtheorem{example}{Example}
\newtheorem{lemma}{Lemma}
\newtheorem{theorem}{Theorem}
\newtheorem{remark}{Remark}

\usepackage{mathtools}
\newcommand\myeq{\stackrel{\mathclap{\normalfont\mbox{$*$}}}{=}}
\begin{document}
	
	\title{A New Hotplug Coded Caching Scheme Using PDAs}
	
\author{Mallikharjuna Chinnapadamala, Charul Rajput and B. Sundar Rajan \\ Department of Electrical Communication Engineering, Indian Institute of Science, Bengaluru \\
	E-mails: chinnapadama@iisc.ac.in, charul.rajput9@gmail.com, bsrajan@iisc.ac.in}
	
	
	
	\maketitle
\begin{abstract}
In the original coded caching model introduced by Maddah-Ali and Niesen in 2014, the server starts broadcasting only after it receives demands from all the users. So, all the users must be active during the delivery phase. In this work, we consider a coded caching model called hotplug coded caching in which some of the users are offline during the delivery phase. This model was first introduced by Ma and Tuninetti (``On Coded Caching Systems with Offline Users," 2022 IEEE International Symposium on Information Theory). The concept of Hotplug Placement Delivery Arrays (HpPDAs) for the hotplug coded caching systems was introduced in (``Improved Hotplug Caching Schemes Using PDAs and $t$-Designs," \emph{arXiv:2311.02856}, 2024), in which the authors have constructed HpPDAs from $t$-designs. This work provides a new hotplug coded caching scheme from the existing HpPDAs. The performance comparison of the proposed scheme with the existing schemes is presented. When applied for HpPDAs from $t$-designs, our scheme outperforms the baseline scheme by Ma and Tuninetti, and the Improved $t$-scheme by Rajput and Rajan in some memory regimes.

\end{abstract}

\section{Introduction}
The concept of coded caching, which was first introduced by Maddah-Ali and Niesen in \cite{MAN}, makes use of locally cached content at the users to reduce the communication load. Coded caching happens in two phases: 1) the Placement Phase and 2) the Delivery Phase. The placement happens during off-peak time. During the placement phase, the server places some files in the cache of each user without knowing the users' future demands. The demands of the users are revealed to the server during the delivery phase. Upon receiving the demands, the server broadcasts the coded multicast messages to satisfy the user's demands. The scheme referred to as the Maddah-Ali and Niesen (MAN) scheme given in \cite{MAN} is proved to be optimal under the assumption of uncoded placement and when the number of files is not less than number of users in \cite{WTP}. When the number of files is less than the number of users, an improved scheme referred to as the YMA (Yu, Maddah-Ali, and Avestimehr) scheme is given in \cite{YMA}. YMA scheme removes some linearly dependent multicast messages that occur when a file is requested by multiple users, thereby reducing the communication load.

The main disadvantage of the MAN scheme is the subpacketization (the number of subfiles each file is divided into) levels. It increases exponentially with the number of users. So, the coded caching schemes with lower subpacketization levels were constructed based on the Placement Delivery Array (PDA) in \cite{YCTC}. PDAs provide placement and delivery mechanisms in a single array. The MAN scheme also corresponds to a class of PDA called MAN PDA \cite{YCTC}. Coded caching schemes from linear block codes and block designs were constructed in \cite{TaR} and \cite{ASK}, respectively, and they have practically achievable subpacketization levels.

In the MAN scheme, the server starts broadcasting only after it receives demands from all the users. So, a limitation of the MAN scheme is that all users must be active and send their demands synchronously. When a certain number of users become inactive during the delivery phase, the server should satisfy the active users. This aspect is considered in another variation of a coded caching model called hotplug coded caching \cite{MaT}. In a hotplug coded caching model, out of $K$ users, only $K'$ users will be active during the delivery phase. During the placement phase, the server knows that $K'$ users will be active but does not know their identity. New coded caching schemes based on Maximum Distance Separable (MDS) coded placement were proposed in \cite{MaT}. We refer to the scheme given in \cite{MaT} as the MT (Ma and Tuninetti) scheme. Hotplug coded caching with demand privacy was studied in \cite{MaT2}. The concept of Placement Delivery Arrays (PDAs) for hotplug coded caching systems was introduced in \cite{RaR}. It provides an algorithm that describes the placement and delivery phase of hotplug coded caching systems using Hotplug PDAs (HpPDAs). Additionally, the construction of a class of HpPDAs using $t$-designs was shown. The construction of a class of HpPDAs based on a class of combinatorial designs called $t$-designs was also shown in \cite{RaR}.

 \subsection{Our Contributions}
In this work, we propose a new algorithm for hotplug coded caching systems based on existing HpPDAs. The proposed scheme outperforms the Improved MT scheme given in \cite{RaR}. However, this improvement comes with an increase in subpacketization. When the proposed algorithm is applied for HpPDAs designed from $t$-designs \cite{RaR}, it outperforms the baseline scheme in \cite{MaT} and Improved $t$-scheme in \cite{RaR} in some memory segments.       

\subsection{Paper Organisation}
In Section \ref{sec2}, the hotplug coded caching problem is defined. Some basic definitions and preliminaries are given in Section \ref{sec3}. Section \ref{sec4} contains the proposed scheme and some examples. The performance comparison of the proposed scheme with the existing schemes is given in Section \ref{sec5}. Finally, Section \ref{sec6} concludes the paper. 

\subsection{Notations}
For a positive integer $n$, $[n]$ denotes the set $\{1,2,\cdots,n\}$. For two sets $A$ and $B$, the notation $A \setminus B$ denotes the set of elements in $A$ which are not in $B$. For a set $A$, the number of elements in it is represented by $|A|$. The binomial coefficient  $\binom{n}{k}$ is equal to $\frac{n!}{k!(n-k)!}$, where $n$ and $k$ are positive integers such that $k \leq n$. For a set $S$ and a positive integer $t$, the notation ${S \choose t}$ represents the collection of all the subsets of $S$ of size $t$.

\section{Hotplug coded caching system}\label{sec2}
In a $(K, K', N)$ hotplug coded caching system, the server stores $N$ files, denoted by $W_{1},W_{2},....W_{N}$ each of size $B$ bits. It is connected to $K$ users through an error-free shared link. Each user is equipped with a cache of size $M$ files. $Z_{k}$ denotes the content of the cache of the user $k\in[K]$. The system has a placement phase and a delivery phase.
\textit{Placement Phase}: The server fills the cache as a function of the files it stores. We assume that the server knows that $K'\leq K $ users will be active during the delivery phase. During the placement phase, the server is unaware of which $K'$ users will be active.\\
\textit{Delivery Phase}: During this phase, the set of active users, denoted by $\mathcal{I} \in \binom{[K]}{K'}$ send their demands. The demand of a user $k$ is denoted by $d_{k}$. Upon knowing the active users and their demands, the server sends the transmissions through the shared link in such a way that the demand of each active user is satisfied using the transmissions and the cached content. The total size of files sent by the server is called the rate, denoted by $R$, of the system.   

Since a $(K,K',N)$ hotplug system cannot perform better than a classical coded caching system with $K'$ users, any converse bound from the classical coded caching system with $K'$ users and $N$ files is applicable for $(K,K',N)$ hotplug system.
The following converse bound was given by Yu et al. in \cite{YMA2018}, which was an improvement of the cutset bound given in \cite{MAN}.
\begin{lemma}[\cite{YMA2018}]\label{bound2}
For $N$ files and $K'$ users each with cache of size $0\leq M\leq N$, the rate $R$ of a classical coded caching system is lower bounded by 
$$R \geq s-1+\alpha- \frac{s(s-1)-\ell(\ell-1)+2\alpha s}{2(N-\ell+1)}M,$$
for any $s \in [\min{(N, K')}], \alpha \in [0, 1]$, where $\ell \in \{1, \ldots, s\}$ is the minimum value such that
$$\frac{s(s-1)-\ell(\ell-1)}{2}+\alpha s \leq (N-\ell+1)\ell.$$
\end{lemma}

\section{Preliminaries} \label{sec3}
In this section, we review the definition of PDAs for dedicated coded caching system and hotplug coded caching system. 
\begin{definition}[Placement Delivery Array \cite{YCTC}]
	For positive integers $K, F, Z$ and $S$, an $F \times K$ array $P = (p_{j,k})_{j \in[F], k \in [K]}$, composed of a specific symbol ``$*$" and $S$ non-negative integers from $[S]$,
	is called a $[K, F, Z, S]$- PDA if it satisfies the following conditions:
	\begin{enumerate}[{C}1.]
		\item The symbol ``$*$" appears $Z$ times in each column,
		\item Each integer occurs at least once in the array,
		\item For any two distinct entries $p_{j_1,k_1}$ and $p_{j_2, k_2}$, $p_{j_1, k_1} =
		p_{j_2, k_2} = s$ is an integer only if
		\begin{enumerate}
			\item $j_1 \neq j_2, k_1 \neq k_2$, i.e., they lie in distinct rows and distinct columns; and
			\item $p_{j_1, k_2} = p_{j_2, k_1} = *$, i.e., the corresponding $2 \times 2$ sub-array formed by rows $j_1, j_2$ and columns $k_1, k_2$ must be of the following form
			$$\begin{bmatrix}
			s & * \\
			* & s
			\end{bmatrix} \quad
			\text{or} \quad
			\begin{bmatrix}
			* & s \\
			s & *
			\end{bmatrix}.$$
		\end{enumerate}
	\end{enumerate}
	
\end{definition}

A $[K, F, Z, S]$-PDA $P$ corresponds to a $(K, N, M)$ coded caching scheme with $K$ users, $N$ files, cache memory of size of $M$ files, subpacketization level $F$, $\frac{M}{N}=\frac{Z}{F}$ and rate $R=\frac{S}{F}$. 

\begin{remark}
	A PDA $P$ is called a $g$-$[K, F, Z, S]$ regular PDA if each integer $s\in [S]$ appears exactly $g$ times in $P$. The MAN scheme corresponds to special class of PDAs called MAN-PDAs. 
\end{remark}
\begin{definition}[Hotplug placement delivery array (HpPDA)]
	Let $K, K', F, F`, Z, Z'$ and $S$ be integers such that $K \geq K', F \geq F'$ and $Z<F'$. Consider two arrays given as follows
	\begin{itemize}
		\item $P=(p_{f,k})_{f \in [F], k \in [K]}$ which is an array containing `$*$' and null. Each column contains $Z$ number of `$*$'s.
		\item $B=(b_{f,k})_{f \in [F'], k \in [K']}$ which is a $[K', F', Z', S]$-PDA.
		
	\end{itemize}
	
	For each $\tau \subseteq [K], |\tau| = K'$, there exists a subset $\zeta \subseteq [F], |\zeta|=F'$ such that 
	\begin{equation}\label{c}
		[P]_{\zeta \times \tau} \myeq B,
	\end{equation}
	where $[P]_{\zeta \times \tau}$ denotes the subarray of $P$ whose rows corresponds to the set $\zeta$ and columns corresponds to the set $\tau$, and $\myeq$ denotes that the positions of all `$*$'s are same in both the arrays. We call it a $(K,K',F, F',Z,Z',S)$- HpPDA $(P, B)$.
\end{definition}

\begin{definition}[Design]
A design is a pair $(X, \mathcal{A})$ such that the following properties are satisfied:
\begin{enumerate}
\item $X$ is a set of elements called points, and
\item $\mathcal{A}$ is a collection of nonempty subsets of $X$ called blocks.
\end{enumerate}
\end{definition}

\begin{definition}[$t$-design]
Let $v, k, \lambda$ and $t$ be positive integers such that $v > k \geq t$. A $t$-$
(v, k, \lambda)$-design is a design $(X, \mathcal{A})$ such that the following properties are satisfied:
\begin{enumerate}
\item $|X| = v$,
\item each block contains exactly $k$ points, and
\item every set of $t$ distinct points is contained in exactly $\lambda$ blocks.
\end{enumerate}
\end{definition}

The following theorem gives the parameters of a hotplug coded caching scheme obtained from an HpPDA.

\begin{theorem}\cite{RaR} \label{thm1}
Given a $(K,K',F, F',Z,Z',S)$-HpPDA $(P, B)$, there exists a $(K, K', N)$ hotplug coded caching scheme obtaining the following memory-rate pair,
$$\left(\frac{M}{N}, R \right)=\left(\frac{Z}{F'}, \frac{S}{F'} \right),$$
where $M$ denotes the number of files each user can store in it's cache and $M<N$.
\end{theorem}

\section{The Proposed Scheme} \label{sec4}
 
 For a given $(K,K',F, F',Z,Z',S)$-HpPDA, we provide a hotplug coded caching scheme, which achieves the following rate memory pair.
 $$\left(\frac{M}{N}, R \right) = \left(\frac{Z}{F'-Z'+Z}, \frac{S}{F'-Z'+Z}\right).$$
 The placement and delivery phases are explained using Algorithm \ref{algo1}. 
Divide each file $W_n, n \in [N]$ into $F'-Z'+Z$ subfiles, and encode these $F'-Z'+Z$ subfiles into $F$ coded subfiles using an $[F, F'-Z'+Z]$ MDS code. For $n \in [N]$, the $F$ coded subfiles of a file $W_n$ are denoted by $C_{n,f}, f \in [F]$. 
 
 \textit{Placement Phase}: In the Algorithm \ref{algo1}, we make use of array $P$ of the HpPDA to fill the cache of each user. It is an $F\times K$ array where columns correspond to users and rows correspond to coded subfiles $C_{n,f}$ of each file $W_{n}, n \in [N]$. The cache of user $k$ contains the subfile $C_{n,f}$ if $p_{f,k}=*$. Since the size of each coded subfile is $1/(F'-Z'+Z)$ and there are $Z$ stars in each column, we have $\frac{M}{N}=\frac{Z}{F'-Z'+Z}$. We encode $F'-Z'+Z$ subfiles of each file $W_n, n \in [N]$, into $F$ coded subfiles using $[F, F'-Z'+Z]$ MDS code with generator matrix $G$ of order $(F'-Z'+Z) \times F$ and fill the caches of all the users as in line 5 in Algorithm 1.

\textit{Delivery Phase}: In the Algorithm \ref{algo1}, we make use of array $B$ for the transmissions. Let $I=\{i_1, i_2, \ldots, i_{K'}\},$ $1 \leq i_k \leq K, \forall k \in [K']$, be the set of active users with demands $D[I]=(d_{i_1}, d_{i_2}, \ldots, d_{i_{K'}})$. By the property of HpPDA, for set $I$ there exists a subset $\zeta \subseteq [F], |\zeta|=F'$ such that $[P]_{\zeta, I} \myeq B$ . With the help of B, we construct a PDA $\overline P$ as given in line 9 of Algorithm \ref{algo1}. Since $\overline{P}$ is a PDA, each integer in $i_k$-th column will provide user $i_k$ a coded subfile of the desired file $d_{i_k}$. There are total $Z'$ stars and $F' - Z'$ integers in a column. As there is a transmission for every integer, each user gets $F'-Z'$ coded subfiles of the desired file from transmissions. Each user already has $Z$ coded subfiles of each file in its cache. So, each user will get $F'-Z'+Z$ coded subfiles of its desired file, which is enough to reconstruct the entire file. The rate achieved is $R=\frac{S}{F'-Z'+Z}$.

 \begin{algorithm}
 	\caption{Hotplug coded caching scheme based on HpPDA $(P, B)$ }
 	
 	\begin{algorithmic}[1]
 		
 		\Function{Placement phase}{$P; W_n, n \in[N]; G$}
 		\State Divide each file $W_n, n \in [N]$ into $F'-Z'+Z$ subfiles, i.e., 
 		$W_n=\{W_{n, f'} | f' \in [F'-Z'+Z]\}.$
 		
 		\State Encode $F'-Z'+Z$ subfiles of each file $W_n, n \in [N]$, into $F$ coded subfiles using $[F, F'-Z'+Z]$ MDS code with generator matrix $G$ of order $(F'-Z'+Z) \times F$, i.e.,
 		$$\begin{bmatrix}
 		C_{n,1} \\ C_{n,2} \\ \vdots \\ C_{n, F} 
 		\end{bmatrix} =  G^{T}\begin{bmatrix}
 		W_{n,1} \\ W_{n,2} \\ \vdots \\ W_{n, F'-Z'+Z} 
 		\end{bmatrix} .$$
 		\For{$k \in [K]$}
 		\State $Z_k \gets \{C_{n,f} \ | \ p_{f,k}=*, n \in [N], f \in [F]\}$.
 		\EndFor
 		\EndFunction 
 		
 		\Function{Delivery phase}{$B; P; W_n, n \in[N]; I; D[I]$}
 		\State Let $I$ be the set of active users with demands $D[I]=(d_{i_1}, d_{i_2}, \ldots, d_{i_{K'}})$. 
 		\State By the property of $(P, B)$ HpPDA, for $I \subseteq [K], |I| = K'$, there exists a subset $\zeta \subseteq [F], |\zeta|=F'$ such that 
 		$[P]_{\zeta \times I} \myeq B.$
 		\State Make a new array $\overline{P}=(\overline{p}_{f,k})_{f \in \zeta, k \in I}$ by filling $s \in [S]$ integers in null spaces of the subarray $[P]_{\zeta \times I}$ in such a way that $\overline{P}=B$.
 		\For{$s \in [S]$}
 		\State Server sends the following coded subfiles:
 		\State $$X_s = \sum_{\overline{p}_{f,k} = s, f \in \zeta, k \in I } C_{d_k, f}.$$
 		\EndFor
 		\EndFunction 
 		
 	\end{algorithmic}
 	\label{algo1}
 \end{algorithm}

 Now we provide some examples of the proposed scheme using two known classes of HpPDAs given in \cite{RaR}: (1) MAN HpPDAs,
 (2) HpPDAs obtained from $t$-designs.
 
 \subsection{MAN HpPDAs}
 For integers $K, K'$ and $t$ such that $K'< K$ and $t\in[K']$, there exists a $(K, K',F, F',Z, Z',S)$ HpPDA $(P, B)$, where 
 \begin{align*}
F'&={K' \choose t}, F={K \choose t}, Z'={K'-1 \choose t-1}, \\
Z&={K-1 \choose t-1}, S={K' \choose t+1}.
\end{align*}
 The array $B$ is a $(K',F',Z',S)$- MAN PDA for $K'$ users. Consider a MAN-PDA $(K,F,Z,S')$ for $K$ users and replace all the integers with null; we can get $P$. When we apply the proposed scheme for the MAN HpPDA, we can recover the scheme mentioned in Remark 2, Appendix B in \cite{MaT}. The following example considers a MAN HpPDA with $K=6, K'=4$ and $t=2$.

\begin{example}\label{ex1}
Consider a $(6,4,15,6,5,3,3)$ HpPDA corresponding to a hotplug coded caching system with $K=6$ users and $K'=4$ active users. For $t=2$, we have $F=15, F'=16, Z=5, Z'=3, S=3$ and the corresponding HpPDA $(P,B)$ is  
\begin{align*}
B&=\begin{bmatrix}
* & * & 1 & 2 \\
* & 1 & * & 3\\
* & 2 & 3 & * \\
1 & * & * & 4\\
2 & * & 4 & * \\
3 & 4 & * & *
\end{bmatrix}, \ 
P=\begin{bmatrix}
* &* & & & &   \\
* &  & * & & & \\
* & &   &* & & \\
* & & & &* & \\
* & & & &  &* \\
 &* & *& & & \\
& * & & * & &\\
& * & & & * & \\
& * & & & & * \\
& & * & * & & \\
& & * & & * & \\
& & * & & & * \\
& & & * & * & \\
& & & * & & * \\
& & & & * & *
\end{bmatrix}.
\end{align*}
Clearly, for each $\tau \subseteq [6], |\tau|=4$, there exists $\zeta =\{S \subseteq \tau \ |\ |S|=2\}, |\zeta|=6$ such that $[P]_{\zeta \times \tau} \myeq B$. Moreover, the rows of array $P$ are denoted by numbers in the set $[15]$ and the columns are denoted by the numbers in the set $[6]$.
Now using HpPDA $(P, B)$ in Algorithm \ref{algo1}, we get a $(6,4,6)$ hotplug coded caching scheme in the following way.

\noindent \textbf{Placement Phase:} Each file is divided into $F'-Z'+Z=8$ subfiles, i.e., $W_n=\{W_{n, 1}, W_{n, 2}, \ldots, W_{n,8}\}, n \in [6]$. Encode all the subfiles of file $W_n$ into $F=15$ coded subfile $C_{n,1}, C_{n,2}, \ldots, C_{n,15}$, using a $[15,8]$ MDS code with generator matrix $G_{8\times 15}$, i.e.,
$$\begin{bmatrix}
C_{n,1} & C_{n,2} & \cdots & C_{n, 15} 
\end{bmatrix} =  \begin{bmatrix}
W_{n,1} & W_{n,2} & \cdots & W_{n, 8} 
\end{bmatrix}G.$$
The caches of all users are filled as follows:
\begin{align*}
Z_1&=\{C_{n,1}, C_{n,2}, C_{n,3}, C_{n,4}, C_{n,5} \ | \ n \in [6]\} \\
Z_2&=\{C_{n,1}, C_{n,6}, C_{n,7}, C_{n,8}, C_{n,9} \ | \ n \in [6]\} \\
Z_3&=\{C_{n,2}, C_{n,6}, C_{n,10}, C_{n,11}, C_{n,12} \ | \ n \in [6]\} \\
Z_4&=\{C_{n,3}, C_{n,7}, C_{n,10}, C_{n,13}, C_{n,14} \ | \ n \in [6]\} \\
Z_5&=\{C_{n,4}, C_{n,8}, C_{n,11}, C_{n,13}, C_{n,15} \ | \ n \in [6]\} \\
Z_6&=\{C_{n,5}, C_{n,9}, C_{n,12}, C_{n,14}, C_{n,15} \ | \ n \in [6]\}. 
\end{align*} 
 
\noindent \textbf{Delivery Phase:} Let the set of active users be $I=\{1,4,5,6\}$ with demands $D[I]=\{2,3,1,5\}$. By constructing a new array $\overline P$ as given in line 11 of Algorithm \ref{algo1}, we get
\begin{align*}
&\begin{matrix}
& 1 & 4 & 5 & 6 
\end{matrix}\\
\overline{P}=  \begin{matrix}
3 \\ 4 \\ 5 \\ 13 \\ 14 \\ 15 
\end{matrix} & \begin{bmatrix}
* & * & 1 & 2 \\
* & 1 & * & 3\\
* & 2 & 3 & * \\
1 & * & * & 4\\
2 & * & 4 & * \\
3 & 4 & * & *
\end{bmatrix}.
\end{align*}

The server transmits the following coded files:
$ X_1= C_{2,13} + C_{3, 4} +C_{1, 3} $,
$ X_2 = C_{2,14} + C_{3, 5} +C_{5, 3} $,      
$ X_3 =   C_{2,15} + C_{1, 5} +C_{5, 4}$,      
$ X_4 = C_{3,15} + C_{1, 14} +C_{5, 13}. $ 
To recover a file, only eight coded subfiles are required because the MDS code used has dimension $8$. The demand of user $1$ is $W_2$. The user $1$ gets $C_{2,13}, C_{2,14}, C_{2,15}$ from $X_1, X_2, X_3$, respectively. Since user $1$ already has $C_{2,1}, C_{2,2}, C_{2,3}, C_{2,4}, C_{2,5}$ in its cache, it can recover file $W_2$. Similarly, user $4,5$ and $6$ will get their desired files.
We have
$$\frac{M}{N}=\frac{Z}{F'-Z'+Z}=\frac{5}{8} \ \text{and} \ R=\frac{S}{F'-Z'+Z}=\frac{1}{2}.$$

\end{example}

 \subsection{HpPDAs obtained from $t$-designs \cite{RaR}}
In this subsection, we briefly explain the construction of HpPDAs from $t$-designs. Let $(X,\mathcal{A})$ be a $t$-$(v,k,\lambda)$ design with non-repeated blocks. From \cite{S2004} it follows that, for  $Y \subseteq X$, where $|Y| = s \leq t$, there are exactly $\lambda_s =\frac{\lambda{v-s \choose t-s}}{{k-s \choose t-s}}$ blocks in $\mathcal{A}$ that contain all the points in $Y$. For $ Y \subseteq T \subseteq X$, where $|T| = t, |Y| = i$ with $i \leq t$, there are exactly
$\lambda_i^{t} =\frac{\lambda{v-t \choose k-i}}{{v-t \choose k-t}}$
blocks in $\mathcal{A}$ that contain all the points in $Y$ and none of the points in $T \backslash Y$. An array $P$ whose columns are indexed by the points in $X$ and rows are indexed by the blocks in $\mathcal{A} $ $(|\mathcal{A}|=b)$ be defined as 
\begin{equation*} \label{arrayP}
P(A,i)=\begin{cases}
* & \text{if} \ \ i \in A, \\
null & \text{if} \ \ i \notin A
\end{cases}.
\end{equation*}
For $0 \leq a_s \leq \lambda_s^{t}$, $1 \leq s \leq t-1,$ consider a set 
$$\mathcal{\mathcal{R}}=\bigcup_{s=1}^{t-1} \left\{ (Y, i) \mid Y \in {[t] \choose s}, i \in [a_s] \right\}.$$ Now, another array $B$ whose columns are indexed by the points in $[t]$ and rows are indexed by the elements in $\mathcal{R}$ is defined as 
\begin{equation*} 
B((Y, i), j)=\begin{cases}
* & \text{if} \ \ j \in Y, \\
\left(Y\cup\{j\},i\right) & \text{if} \ \ j \notin Y
\end{cases}.
\end{equation*}
The pair $(P,B)$ forms a $(K,K',F,F',Z,Z',S)$-HpPDA, where
\begin{align*}\label{parameters1}
 \nonumber &K=v, K'=t, F=b, F'=|\mathcal{R}|=\sum_{s=1}^{t-1} a_s {t \choose s}, 
Z=\lambda_1,\\ & Z'=\sum_{s=1}^{t-1} a_s {t-1 \choose s-1}, 
 \text{and} \ S=\sum_{s=1}^{t-1} a_s {t \choose s+1},   
\end{align*}
 
 \begin{example}\label{ex2}
     Consider an HpPDA designed from $t$-design in \cite{RaR}. The following HpPDA is a $(8,3,14,9,7,5,5)$ HpPDA that corresponds to $(8,3,N)$- hotplug coded caching system

     \[
P=\begin{blockarray}{ccccccccc}
& 1 & 2 & 3 & 4 & 5 & 6 & 7 & 8 \\
\begin{block}{c[cccccccc ]}
1 &* & * &  &  & * & * & & \\
 2 & &  & * & * &  &  & * & *\\
 3 & & * &  & * &  & * &  & *\\
4 & * &  & * &  & * &  & * & \\
5 & *&  &  & * & * &  &  & *\\
6 & & * & * &  &  &*  & * & \\
7 & *& * & * & * &  &  &  & \\
 8 & &  &  &  & * & * & * & *\\
9 & *& * &  &  &  &  & * & *\\
10 & &  & * & * & * &*  &  & \\
11 & *&  &*  &  &  &*  &  & *\\
12 & & * &  &*  & * &  & * & \\
13 & *&  &  & * &  & * & * & \\
14 & & * & * &  & * &  &  & *\\
\end{block}
\end{blockarray},
B=
\begin{blockarray}{ccc}
 & &  \\
\begin{block}{[ccc]}
* & * &  1 \\
  * & 1 & *\\
 1 & * & *\\
* & * & 2  \\
 * &2 & *\\
  2 & * & *\\
* & 3 &  4 \\
 3 & * & 5 \\
4 & 5 & *\\
\end{block}
\end{blockarray}.
\]

\noindent \textbf{Placement Phase:} Each file is divided into $F'-Z'+Z=11$ subfiles, i.e., $W_n=\{W_{n, 1}, W_{n, 2}, \ldots, W_{n,11}\}, n \in [6]$. Encode all the subfiles of file $W_n$ into $F=14$ coded subfile $C_{n,1}, C_{n,2}, \ldots, C_{n,14}$, using a $[14,11]$ MDS code with generator matrix $G_{11\times 14}$, i.e.,
$$\begin{bmatrix}
C_{n,1} & C_{n,2} & \cdots & C_{n, 14} 
\end{bmatrix} =  \begin{bmatrix}
W_{n,1} & W_{n,2} & \cdots & W_{n, 11} 
\end{bmatrix}G.$$
The caches of all users are filled as follows:
\begin{align*}
Z_1&=\{C_{n,1}, C_{n,4}, C_{n,5}, C_{n,7}, C_{n,9}, C_{n,11}, C_{n,13} \ | \ n \in [6]\} \\
Z_2&=\{C_{n,1}, C_{n,3}, C_{n,6}, C_{n,7}, C_{n,9}, C_{n,12}, C_{n,14} \ | \ n \in [6]\} \\
Z_3&=\{C_{n,2}, C_{n,4}, C_{n,6}, C_{n,7}, C_{n,10}, C_{n,11}, C_{n,14} \ | \ n \in [6]\} \\
Z_4&=\{C_{n,2}, C_{n,3}, C_{n,5}, C_{n,7}, C_{n,10}, C_{n,12}, C_{n,13} \ | \ n \in [6]\} \\
Z_5&=\{C_{n,1}, C_{n,4}, C_{n,5}, C_{n,8}, C_{n,10}, C_{n,12}, C_{n,14} \ | \ n \in [6]\} \\
Z_6&=\{C_{n,1}, C_{n,3}, C_{n,6}, C_{n,8}, C_{n,10}, C_{n,11}, C_{n,13} \ | \ n \in [6]\}. \\
Z_7&=\{C_{n,2}, C_{n,4}, C_{n,6}, C_{n,8}, C_{n,9}, C_{n,12}, C_{n,13} \ | \ n \in [6]\}. \\
Z_8&=\{C_{n,2}, C_{n,3}, C_{n,5}, C_{n,8}, C_{n,9}, C_{n,11}, C_{n,14} \ | \ n \in [6]\}. \\
\end{align*} 
\noindent \textbf{Delivery Phase:} Let the set of active users be $I=\{2,6,8\}$ with demands $D[I]=\{1,3,4\}$. By constructing a new array $\overline P$ as given in line 11 
Algorithm \ref{algo1}, we get

\[
\overline P =\begin{blockarray}{cccc}
& 2 &   6  & 8 \\
\begin{block}{c[ccc ]}
1&* & * & 1\\
9 & *& 1 &  *\\
8 &1 & * & *\\
6 &* & * &2 \\
14 &* &2 &* \\
11 &2 & * & *\\
7 & *  &3  &4 \\
10  &3 & * &5  \\
 2 &4 &5  & *\\
\end{block}
\end{blockarray}.
 \]
The server transmits the following coded files:
$
X_1= C_{1,8} + C_{3, 9} +C_{4, 1}$,
$X_2 = C_{1,11} + C_{3, 14} +C_{4, 6}$,
$X_3 =   C_{1,10} + C_{3, 7}$,
$X_4 = C_{1,2} + C_{4, 7}$,
$X_5 = C_{3,2} + C_{4, 10}$.
To recover a file, only eleven coded subfiles are required because the MDS code used has dimension $11$. Consider user 2. The demand of the user $2$ is $W_1$. The user $2$ gets $C_{1,8}, C_{1,11}, C_{1,10}, C_{1,2}$ from $X_1, X_2, X_3, X_4 $, respectively. Since user $2$ already has $C_{1,1}, C_{1,3}, C_{1,6}, C_{1,7}, C_{1,9}, C_{1,12}, C_{1,14} $ in its cache, it can recover file $W_1$. Similarly, users $6$ and $8$ will get their desired files.
We have
$$\frac{M}{N}=\frac{Z}{F'-Z'+Z}=\frac{7}{11} \ \text{and} \ R=\frac{S}{F'-Z'+Z}=\frac{5}{11}.$$

 \end{example}

 \section{Comparison} \label{sec5}
In this section, we compare the proposed scheme with the other schemes in the literature. One HpPDA corresponds to one memory point for a $(K, K',N)$ hotplug coded caching system. However, corresponding to each $t \in [K']$, we get a different MAN HpPDA, which corresponds to a different memory point. Therefore, we get a scheme for $K'$ memory points for a $(K, K', N)$ hotplug coded caching system. Similarly, corresponding to one $t$-design, we get multiple HpPDAs using the $t$-scheme given in \cite{RaR}. Each HpPDA constructed using a $t$-design corresponds to the same hotplug coded caching system with different memory sizes. Using these multiple memory points, we compare the performance of the proposed scheme with the existing schemes.

 \begin{figure*} 
     \centering
     \includegraphics[width=\linewidth]{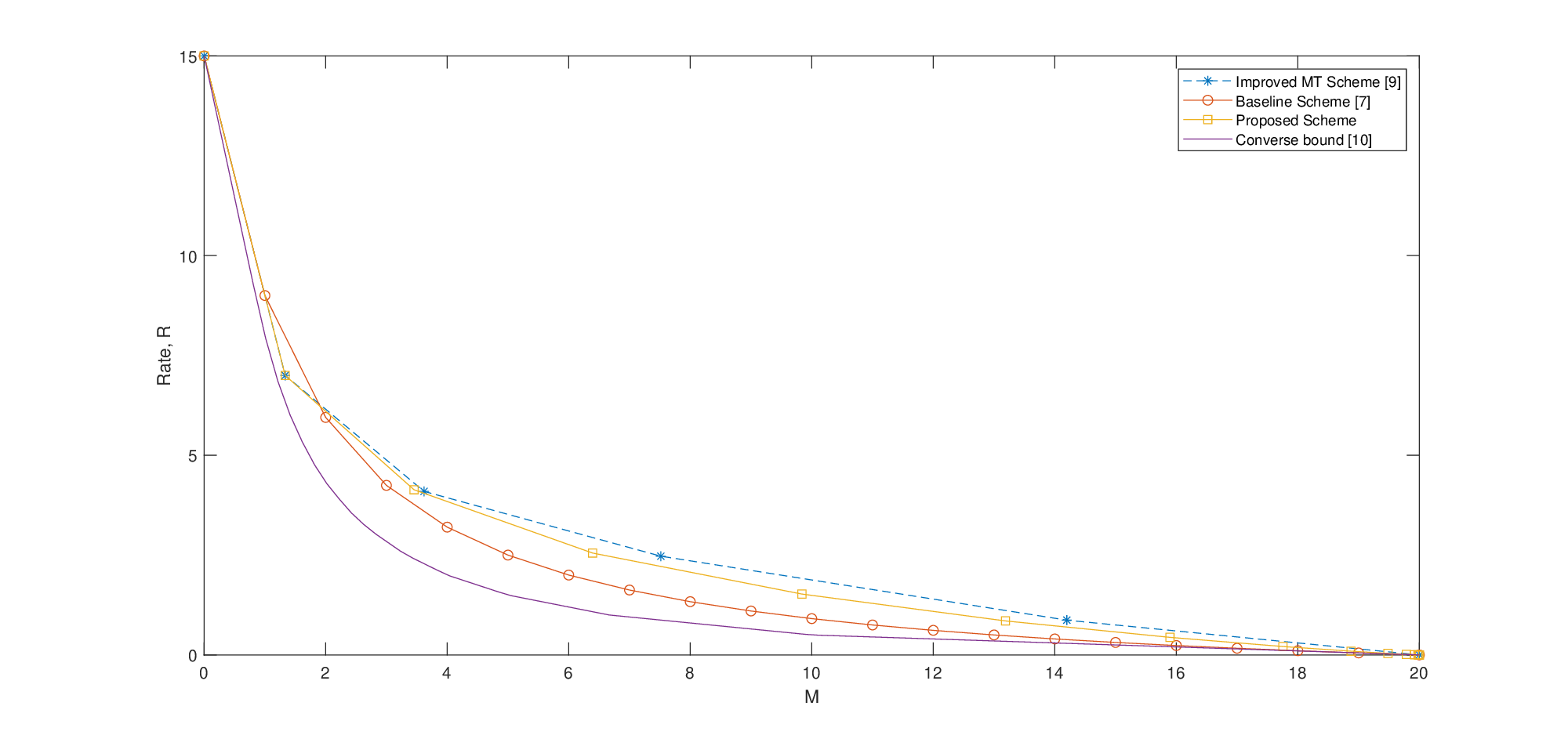}
     \caption{$(20, 15, 20)$ Hotplug coded caching system.}
     \label{Fig1}
 \end{figure*}

For $K=20$ and $K'=15$, we get a $\left(20, 15, {20 \choose{t}}, {15 \choose{t}}, {19 \choose{t-1}}, {14 \choose{t-1}}, {15 \choose{t+1}} \right) $ MAN HpPDA for all $t\in [15]$. Using these HpPDAs, we compare the performance of all the schemes for a $(20,15,20)$ hotplug coded caching system in Figure \ref{Fig1}. Clearly, the proposed scheme outperforms the Improved MT scheme in \cite{RaR}. However, the Baseline scheme in \cite{MaT} performs better than the proposed scheme. 

 \begin{figure*}
     \centering
     \includegraphics[width=\linewidth]{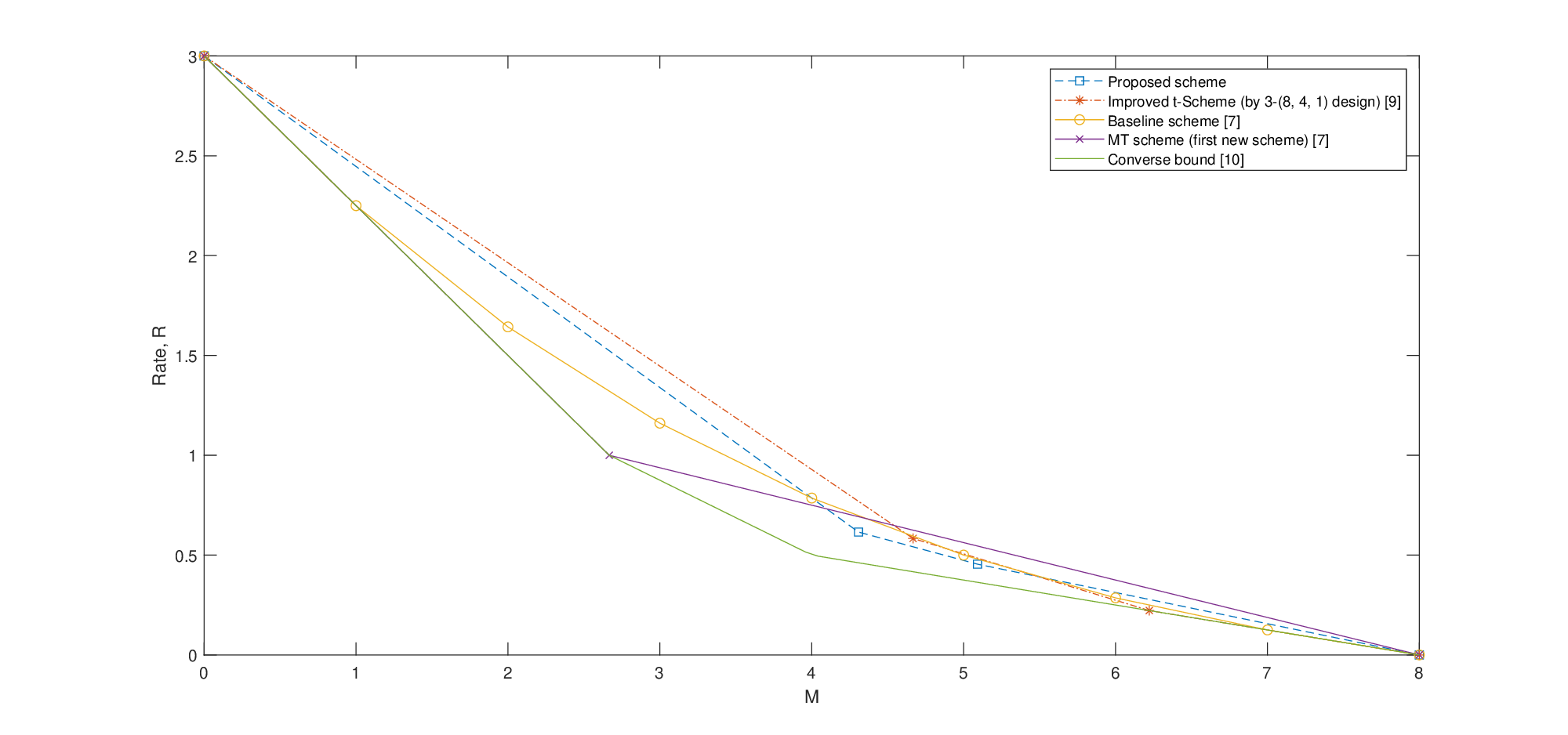}
     \caption{$(8, 3, 8)$ Hotplug coded caching system.}
     \label{Fig2}
 \end{figure*}

When the proposed scheme is applied for HpPDAs obtained from $t$-designs, it performs better than the baseline scheme in \cite{MaT} and Improved $t$-scheme in \cite{RaR} in some memory segments. In Fig. \ref{Fig2}, we consider $(8, 3, 8)$ hotplug coded caching system. From a $3$-$(8, 4, 1)$ design, we get two HpPDAs \cite{RaR} with the following parameters
\begin{enumerate}
\item a $(8, 3, 14, 9, 7, 5, 5)$ HpPDA (given in Example \ref{ex2}),
\item a $(8, 3, 14, 12, 7, 6, 8)$ HpPDA. 
\end{enumerate}
From Theorem \ref{thm1}, using these HpPDAs in the hotplug coded caching scheme given in \cite{RaR}, we get the memory-rate points $\left(\frac{7}{9}, \frac{5}{9}\right) \ \text{and} \ \left(\frac{7}{12}, \frac{8}{12}\right)$, respectively. Further by Improved $t$-scheme given in \cite{RaR}, these memory-rate points were improved to $\left(\frac{7}{9}, \frac{2}{9}\right) \ \text{and} \ \left(\frac{7}{12}, \frac{7}{12}\right)$, respectively. In Fig. \ref{Fig2}, these points are represented by '*' and the rate memory tradeoff attained by the Improved $t$-scheme is shown by a red dashed-dotted line.

Using the same HpPDAs obtained from a $3$-$(8, 4, 1)$ design in the proposed scheme, we get the following memory-rate points 
$$\left(\frac{7}{11}, \frac{5}{11}\right) \ \text{and} \ \left(\frac{7}{13}, \frac{8}{13}\right),$$
which are shown by blue squares in Fig. 2, and the proposed scheme is shown by a blue dashed line.

Further, the MT scheme, which is referred to as \textit{first new scheme} in \cite{MaT}, and the baseline line scheme \cite{MaT}, which is nothing but the MAN scheme with the restricted demand set, are plotted in Fig. \ref{Fig2} for the $(8,3,8)$ hotplug coded caching system. The converse bound given in Lemma \ref{bound2} is shown by a green line in Fig. \ref{Fig2}.

As can be seen in Fig. \ref{Fig2}, both the memory-rate points attained by the proposed scheme beat other schemes, which lead to the improvement in the memory range (4.0975, 5.4902). It is clear from Theorem \ref{thm1} that for a given $(K, K', F, F', Z, Z', S)$ HpPDA, the scheme given in \cite{RaR} works for cache memory $M=\frac{Z}{F'}$, whereas for the same HpPDA, the proposed scheme  works for cache memory $M=\frac{Z}{F'-Z'+Z}$. As $Z' \leq Z$, we have $\frac{Z}{F'-Z'+Z}\leq \frac{Z}{F'}$. Therefore, using the proposed scheme for an HpPDA, we get a scheme for a lesser cache memory size. For example, using the HpPDAs obtained from $3$-$(8,4,1)$ design, we get the cache memories $\frac{7}{12}$ and $\frac{7}{9}$ in scheme given in \cite{RaR}, whereas in the proposed scheme, we get the cache memories $\frac{7}{13}$ and $\frac{7}{11}$, respectively. Therefore, in lower memory range, the proposed scheme performs better than the scheme given in \cite{RaR} using an HpPDA.

  \begin{figure*}
     \centering
     \includegraphics[width=\linewidth]{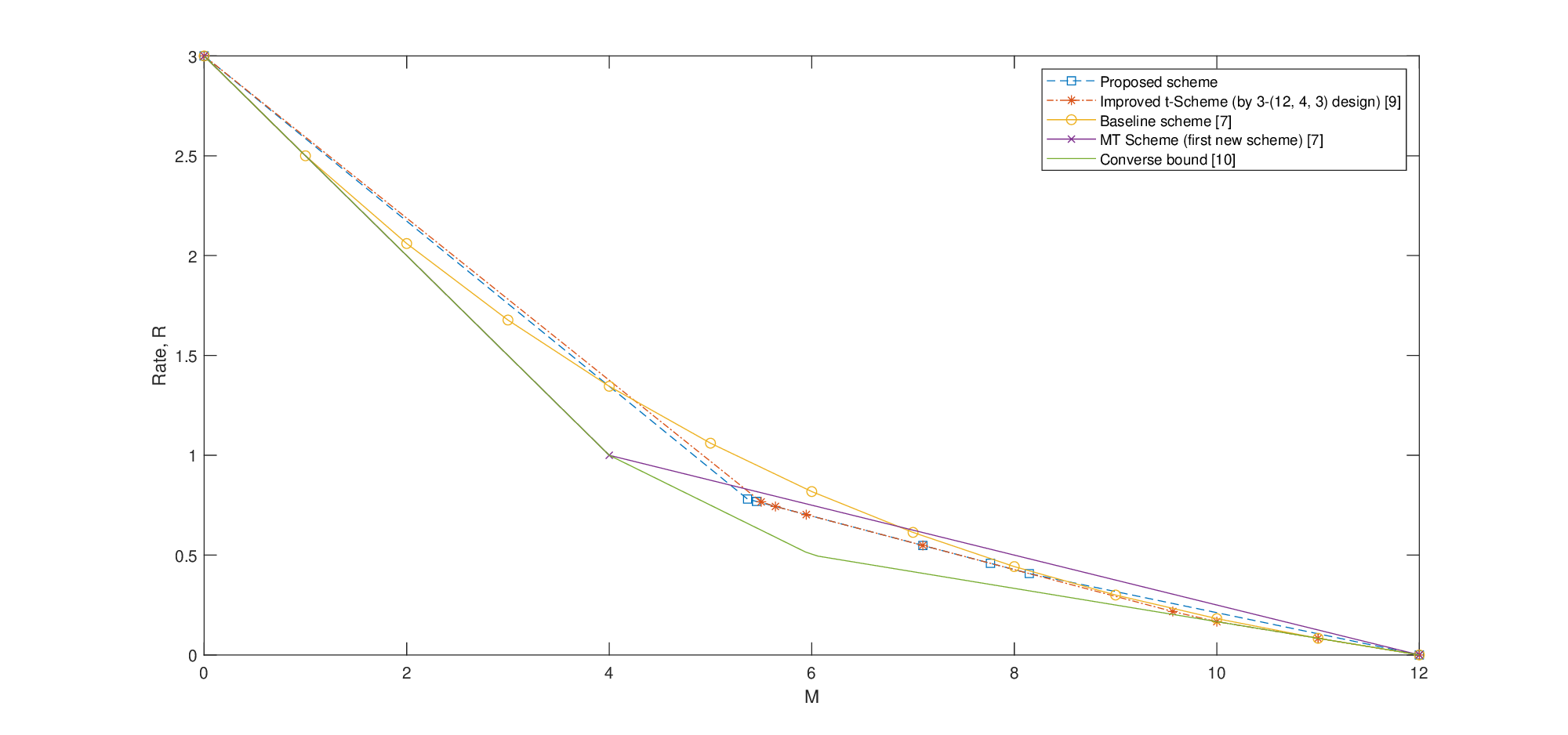}
     \caption{$(12, 3, 12)$ Hotplug coded caching system.}
     \label{Fig3}
 \end{figure*}
 \begin{figure*}
     \centering
     \includegraphics[width=\linewidth]{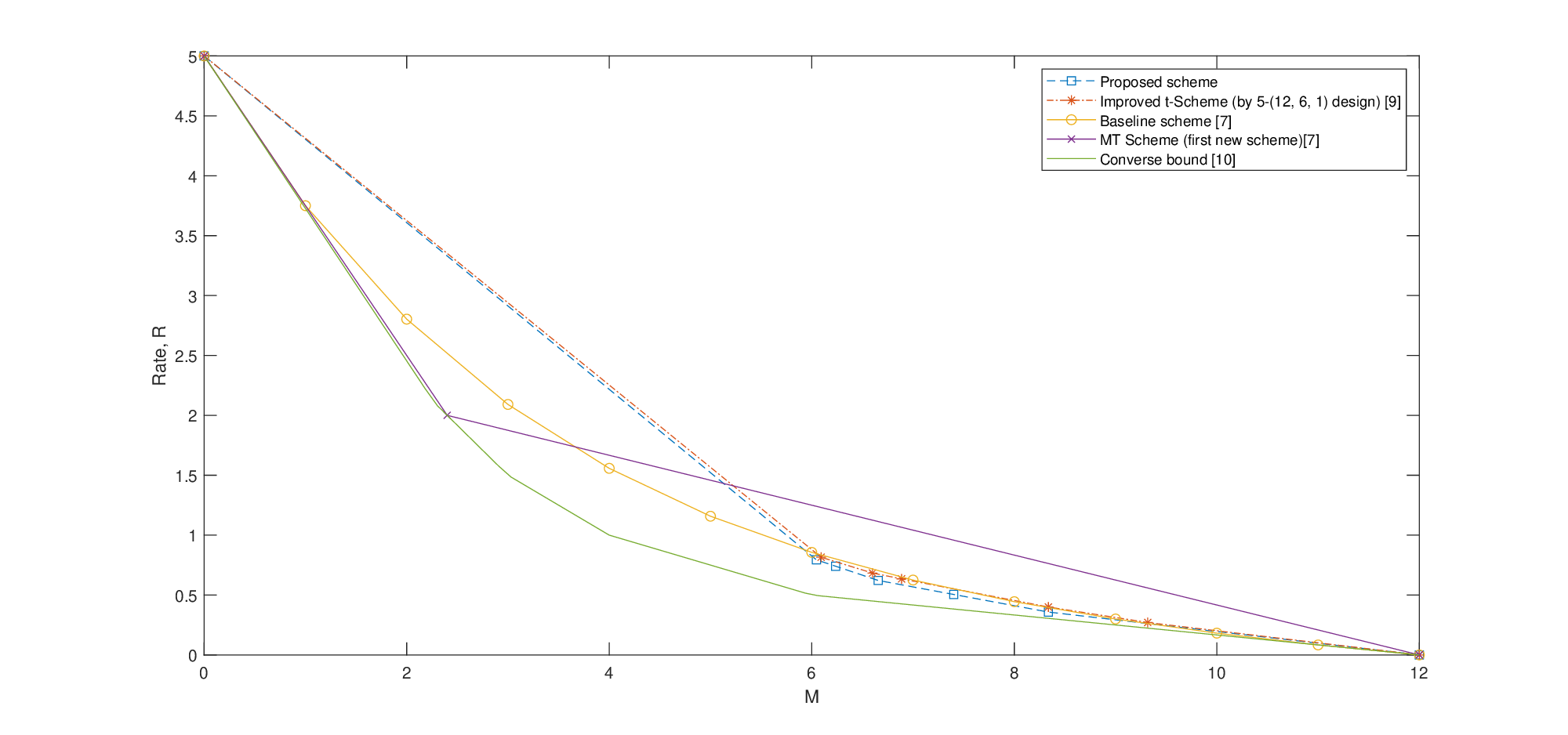}
     \caption{$(12, 5, 12)$ Hotplug coded caching system.}
     \label{Fig4}
 \end{figure*}

In a similar manner, Fig. \ref{Fig3} and Fig. \ref{Fig4} demonstrate the comparison of rate-memory tradeoffs attained by the proposed scheme with the rate-memory tradeoffs attained by the other schemes. Further, the cache memory range for which the proposed scheme gives a better rate than the other schemes is given in Table \ref{tab1} for Fig. \ref{Fig1}, Fig. \ref{Fig3} and Fig. \ref{Fig4}.

\begin{table}[!t]
\caption{Cache memory ranges where the proposed scheme is better than the other schemes. \label{tab1}}
\centering
\begin{tabular}{|c|c|c|c|c|c|} 
 \hline
   & Hotplug system & \multicolumn{2}{c|}{From}
            & \multicolumn{2}{c|}{To} \\ 
 \hline
& $(K, K', N)$ & $M$ & $R$ & $M$ & $R$ \\
\hline
Fig. \ref{Fig2} & $(8,3,8)$ &$4.0975$ & $0.7317$ & $5.4902$ & $0.3922$\\

Fig. \ref{Fig3} & $(12,3,12)$ & $5.1968$ & $0.8504$ & $5.64103$ & $0.74359$\\
Fig. \ref{Fig4} & $(12, 5, 12)$ & $5.9231$ & $0.8793$ & $12$ & $0$\\
\hline
\end{tabular}
\end{table}

 \section{Conclusion} \label{sec6}
 
 For the hotplug coded caching system, we proposed a new scheme based on existing HpPDAs. With an increase in subpacketization, the proposed scheme performs better than the Improved MT scheme in \cite{RaR}. Additionally, it was shown for multiple examples that the proposed scheme outperforms the baseline scheme in \cite{MaT} and Improved $t$-scheme in \cite{RaR} in some memory segments.

\section*{Acknowledgments}
This work was supported partly by the Science and Engineering Research Board (SERB) of the Department of Science and Technology (DST), Government of India, through J.C. Bose National Fellowship to Prof. B. Sundar Rajan and by the C V Raman postdoctoral fellowship awarded to Charul Rajput.

\end{document}